\documentclass[sigplan,10pt,nonacm=true,authorversion=true]{acmart}

\newif\ifsubmission
\submissiontrue

\usepackage{subcaption}
\usepackage[english]{babel}
\usepackage{graphicx}
\usepackage{multirow}
\usepackage{xcolor}
\usepackage{amsmath}
\usepackage{array}

\usepackage{url}

\usepackage[font=small,format=plain,labelfont=bf,textfont=bf]{caption}

\usepackage{colortbl}
\usepackage{float}
\usepackage{etoolbox}

\usepackage{booktabs}
\usepackage{xspace}
\usepackage{enumitem}
\usepackage{textcomp}
\usepackage{balance}
\usepackage{xpatch}

\newcommand{\smartparagraph}[1]{\noindent{\bf #1}\ }

\begin{document}

\newfont{\thisttlfnt}{phvb8t at 20pt}
\newfont{\thissttlfnt}{phvb8t at 16pt}

\pagestyle{plain}

\title{With Great Freedom Comes Great Opportunity:\\
{\huge Rethinking Resource Allocation for Serverless Functions}}

\author{Muhammad Bilal}
\authornote{Work done in part while author was interning at KAUST.}
\affiliation{%
  \institution{UCLouvain and IST(ULisboa)/INESC-ID}
}

\author{Marco Canini}
\affiliation{%
  \institution{KAUST}
}

\author{Rodrigo Fonseca}
\affiliation{%
  \institution{Microsoft Research}
}

\author{Rodrigo Rodrigues}
\affiliation{%
  \institution{IST(ULisboa)/INESC-ID}
}

\begin{abstract}
Current serverless offerings give users a limited degree of flexibility for configuring the resources allocated to their function invocations by either coupling memory and CPU resources together or providing no knobs at all. These configuration choices simplify resource allocation decisions on behalf of users, but at the same time, create deployments that are resource inefficient. 

In this paper, we take a principled approach to the problem of resource allocation for serverless functions, allowing this choice to be made in an automatic way that leads to the best combination of performance and cost. In particular, we systematically explore the opportunities that come with decoupling memory and CPU resource allocations and also enabling the use of different VM types. 
We find a rich trade-off space between performance and cost. The provider can use this in a number of ways: from exposing all these parameters to the user, to eliciting preferences for performance and cost from users, or by simply offering the same performance with lower cost. This flexibility can also enable the provider to optimize its resource utilization and enable a cost-effective service with predictable performance.

 Our results show that, by decoupling memory and CPU allocation, there is potential to have up to 40\% lower execution cost than the preset coupled configurations that are the norm in current serverless offerings. Similarly, making the correct choice of VM instance type can provide up to 50\% better execution time. Furthermore, we demonstrate that providers can utilize different instance types for the same functions to maximize resource utilization while providing performance within 10-20\% of the best resource configuration for each respective function.  
 
\end{abstract}

\maketitle

\section{Introduction}
The serverless programming model has flourished in the last few years, mainly because it allows developers to concentrate on the application logic and not worry about scalability and resource management. Developers only have to write the code for one or more cloud functions, and have little to no control on the amount of resources the cloud provider uses to run them. 
Cloud providers take care of provisioning, deployment, scalability, and maintenance of the resources required for each function invocation. 

Current serverless offerings in the cloud typically \emph{couple} memory and CPU resource allocations together. Both AWS~\cite{lambda} and GCP~\cite{gcp} provide preset resource allocation configurations. In particular, AWS assigns a CPU share proportional to the amount of memory requested by the user (in a fine-grained way, but up to 10 GB); GCP provides users with seven preset resource allocation options to choose from.
Azure Functions~\cite{azurefunctions}, in turn, guarantees at least 1 vCPU core to each function instance and allows up to 1.5GB of memory per function instance (but the user is charged based on the actual memory consumption).  
Additionally, none of the cloud providers provide the option to select the VM type that runs the function, even though the VM type used to run serverless functions is not always the same~\cite{wang2018peeking, maissen2020faasdom}. 

This simple interface is one of the defining characteristics of serverless computing, with the advantage of removing the configuration burden from the user. This is in stark contrast with the complexity of selecting and provisioning machines from the \emph{hundreds} of configuration options and VM types in regular IaaS offerings~\cite{yadwadkar2017selecting,alipourfard17cherrypick}.
However, it also comes with significant disadvantages.

First, as we show later, in most cases, there are configurations that achieve better performance, better cost, or both, than the ones from current offerings.
Second, when there are knobs, they are low-level: it is difficult for most users to translate memory and CPU to performance and cost.
Third, the lack of full transparency hurts predictability, and even raises the question of whether the cloud provider is making the right choices to optimize resource usage, translating into lower costs for the users. 

In this paper, we take a step back and rethink, from a clean slate, the allocation of resources for the execution of serverless functions. In particular, we try to determine what can be gained by taking fine-grained control over the individual allocation of CPU, memory, and VM type to each serverless invocation. This requires us to understand how these allocation decisions influence the trade-off between performance and monetary cost, the predictability of these metrics, and ability to meet target execution times. Furthermore, we want to understand what is the minimal resource allocation that is required to meet such targets, if possible leveraging idle resources, whose type and availability may vary with time.

Despite the benefits of flexibility, simply offering a much larger configuration space to users negates the advantages of simplicity. Therefore, we also 
 provide a thorough study of the effectiveness of  \emph{black-box} optimization algorithms to automatically determine the right resource configuration to remove the need for the user to deal with the complexity of choice, or to profile their functions to determine the right resources. The output of these algorithms can then be used directly in two possible ways: 
either by a cloud provider to \emph{automatically} allocate resources for a user's functions, or by the serverless end-user if the cloud provider provides the \emph{right configuration knobs}. We discuss different interfaces that an auto-tuning framework built using black-box optimization algorithms can provide to the end-user, achieving the best of both worlds: maintaining simplicity, while giving the user control of both cost and execution time.
Alternatively, if automatic resource configuration is provided by the cloud vendor, we explore whether cloud providers can allocate the functions on different VM types to maximize the utilization of spare resources while providing predictable performance.

Our evaluation shows that the more flexible resource allocation option can provide up to 40\% better execution time and 50\% better execution cost than the restrictive resource allocation strategy that cloud providers currently use. We show that black-box optimization algorithms can reach within 10\% of the performance of the best configuration in our resource allocation search space within 20 optimization trials.
Finally, we show that cloud providers can reduce their costs (or conversely increase resource utilization) by utilizing different instance types for the same function, while providing performance within 10\% of the best configuration.

\smartparagraph{Contributions.}
We make the following contributions:\\
\noindent $\bullet$ We determine the ground truth about the execution time and cost of 6 serverless applications across 288 resource configurations and multiple inputs (\S\ref{sec:motivation}).\\
\noindent $\bullet$ We analyze the potential benefits of enabling a more flexible resource allocation for serverless functions (\S\ref{sec:opportunity}).\\
\noindent $\bullet$ We analyze the accuracy of 4 Bayesian Optimization algorithms for determining the best resource allocation in terms of execution time and execution cost (\S\ref{sec:auto-opt}).\\
\noindent $\bullet$ We determine whether the serverless functions of our study have data-dependent performance characteristics  (\S\ref{sec:input-dep}).\\
\noindent $\bullet$ We propose a set of possible interfaces for enabling the user to benefit from multi-objective optimization  (\S\ref{sec:mo-interface}).\\
\noindent $\bullet$ We evaluate the cost reduction opportunities for the cloud providers by using different instance types while providing predictable performance  (\S\ref{sec:different-instance-types}). 

\section{Motivation}
\label{sec:motivation}
Resource allocation decisions for functions can significantly impact the performance and cost of serverless functions. To characterize this performance and execution cost variation, we
exhaustively test six benchmark functions on the resource configuration space defined in Table~\ref{tab:search-space}.
(We defer to \S\ref{sec:setup} further details of the setup for these experiments, including a description of the functions.)
We run these benchmarks on AWS and adopt their nomenclature of instance families.
As AWS does not provide information about the per-core or per-GB price, we calculate the execution cost based on assumptions and methodology defined in \S\ref{sec:cost-calculation}.

A resource allocation choice specifies a CPU share, a memory limit, and the instance family. 
The CPU share is the timeshare of a vCPU allocated to the function.
So, a share of 0.25 means that a quarter of vCPU time is allocated to the function whereas a share of 1 or 2 implies allocations of 1 and 2 entire vCPUs, respectively. The memory limit, expressed in MB, is simply the amount of memory allocated to the function. In addition, we run functions on a diverse range of instance families, which influence the CPU type used by the function.
The caption of Table~\ref{tab:search-space} lists their nomenclature.

\begin{table}[]
\small
\begin{tabular}{ll}
\toprule
\multicolumn{1}{c}{\textbf{Resource Configuration}} & \multicolumn{1}{c}{\textbf{Values}} \\ \midrule 
CPU share                                    & {[}0.25, 0.5, 0.75, 1, 1.25, 1.5, 1.75, 2{]}          \\ 
Memory limit (MB)                           & {[}128, 256, 512, 768, 1024, 2048{]} \\ 
Instance families                                & {[}c6g, m6g, c5, m5, c5a, m5a{]}               \\ \bottomrule
\end{tabular}
\caption{Resource allocation search space. A configuration is a value for CPU share, memory limit and instance family. For instance family names, a prefix `c' means compute-optimized, and `m' is a general-purpose instance; a suffix `g' means Graviton2 ARM-based, `a' AMD-based, and no suffix corresponds to Intel-based instances.
The search space has 288 configurations.}
\label{tab:search-space}
\end{table}

\begin{figure}
    \centering
    \includegraphics[width=\linewidth]{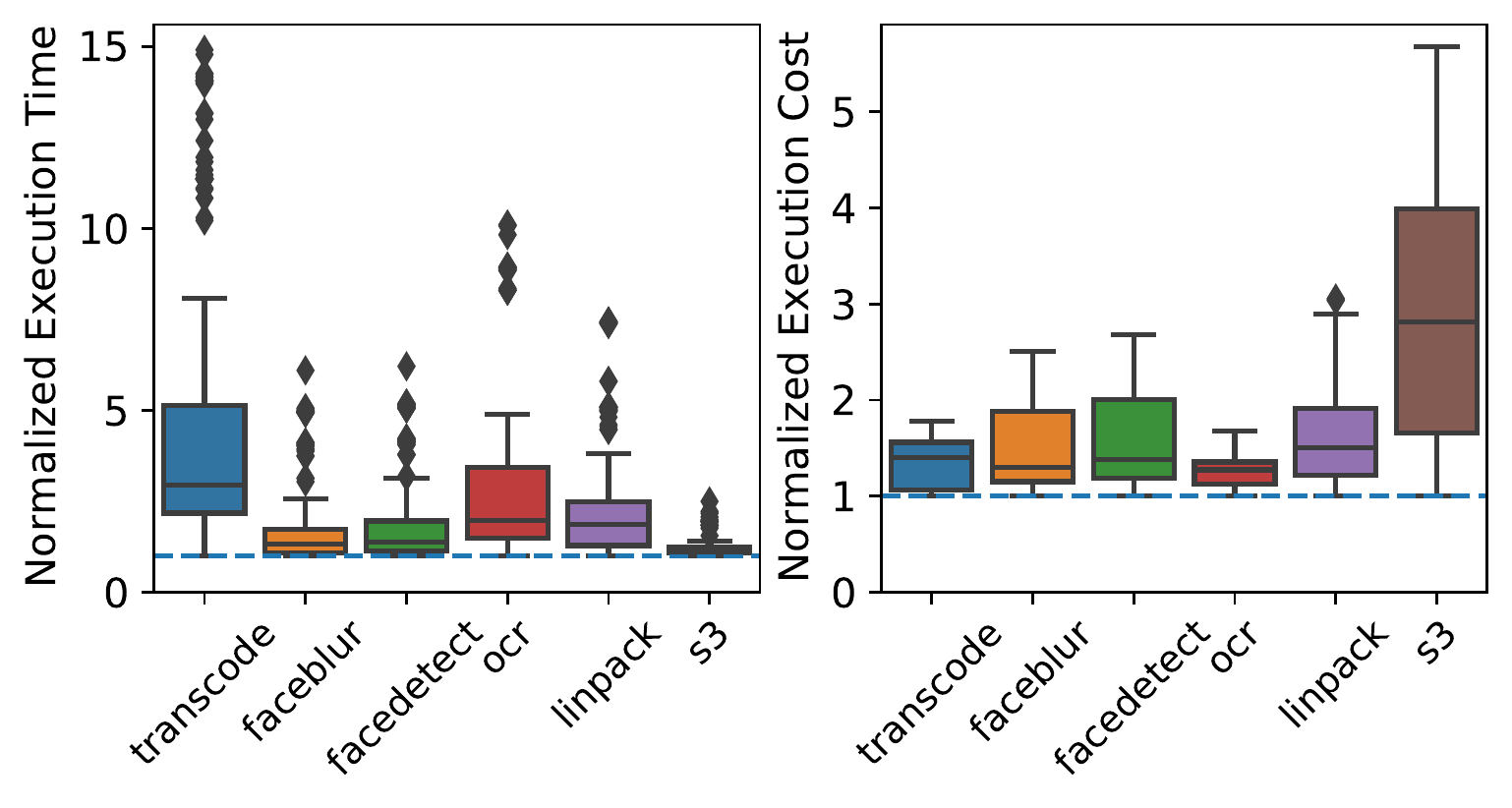}
\caption{Execution time and cost of each function across the entire configuration space defined in Table~\ref{tab:search-space}, normalized w.r.t.\ the best configuration of each function.}
\label{fig:motivation}
\end{figure}

Figure~\ref{fig:motivation} shows a box plot summary\footnote{All boxplots in the paper show median, 1st and 3rd quartiles with whiskers showing the distribution 1.5$\times$ IQR past the high and low quartiles and anything beyond is shown as outliers.} 
of the normalized execution time and execution cost for each function across the entire configuration search space of Table~\ref{tab:search-space}. For each function, the normalization is done w.r.t. the best (minimum) execution time and execution cost for that function in the search space.

We observe that in the worst case, selecting the wrong configuration can lead up to 14.9$\times$ worse execution time and 5.6$\times$ worse execution cost compared to the best configuration. From this result, we conclude that an incorrect resource allocation can lead to significant performance and cost penalties.

\section{Experimental setup}
\label{sec:setup}
We use the OpenFaaS~\cite{openfaas} serverless framework to execute and measure performance of a diverse range of benchmark functions across the resource allocation search space. We deploy OpenFaaS atop the k3s distribution~\cite{k3s} of Kubernetes as a cluster running on AWS EC2 instances. We prepare multi-architecture Docker containers (amd64 and arm64) for all benchmark functions using \textit{docker buildx} without using any platform-specific optimized libraries.

We collect the ground truth performance and cost data as follows.
We execute each function using multiple input data samples on every resource allocation in the configuration search space of Table~\ref{tab:search-space}.
To minimize the impact of performance outliers, we execute each function at least 5 times on a given configuration. We use a function execution timeout of 600s, which is comparable to the timeouts in current serverless offerings~\cite{wen2020understanding}. The input samples -- while arbitrary -- were chosen from publicly available datasets.
One input sample is used as the default but we analyze how performance depends on input data (a modest 20\% at most, \S\ref{sec:input-dep}).
Overall, we run over 5,000 combinations of resource configurations, benchmarks and input samples.
From this ground truth data, we identify the overall best configuration for each function with regard to both execution time and execution cost.

\subsection{Benchmark functions}
\label{sec:benchmarks}

Table~\ref{tab:benchmarks} shows the benchmark functions used in this work. These benchmarks are a mix of applications taken from prior works and applications from OpenFaaS function store~\cite{openfaas-store}.
{\sf transcode}'s and {\sf ocr}'s request handlers are in Python but use internal bindings to invoke C and C++ applications, respectively. {\sf facedetect} and {\sf faceblur} use Go libraries~\cite{pigo, stackblur-go}. {\sf linpack} uses an application from FunctionBench~\cite{functionbench-github}. Both {\sf transcode} and {\sf ocr} are able to effectively utilize $> 1$ vCPU.

\begin{table}[]
\resizebox{0.48\textwidth}{!}{%
\begin{tabular}{llll*{4}{>{\centering\arraybackslash}p{1.5em}}}
\toprule
\multicolumn{1}{c}{\multirow{2}{*}{\textbf{Name}}} & \multicolumn{1}{c}{\multirow{2}{*}{\textbf{Purpose}}}                                                          & \multirow{2}{*}{\textbf{Language}} & \multicolumn{4}{c}{\textbf{Important Resources}}                                                                                     \\ \cmidrule{4-7} 
\multicolumn{1}{c}{}                               & \multicolumn{1}{c}{}                                                                                           &                                    & \multicolumn{1}{c}{\textbf{C}} & \multicolumn{1}{c}{\textbf{P}} & \multicolumn{1}{c}{\textbf{M}} & \multicolumn{1}{c}{\textbf{N}} \\ \midrule
{\sf facedetect}                                           & Image face detection                                                                                            & Go                                 &     $\bullet$                           &                                 &                                 &                                 \\ 
{\sf faceblur}                                             & Image face blurring                                                                                             & Go                                 &     $\bullet$                            &                                 &                                 &                                 \\ 
{\sf transcode}                                            & Video transcoding                                                                                               & Python \& C                        &       $\bullet$                          &     $\bullet$                            &     $\bullet$                            &                                 \\
{\sf ocr}                                                  & Optical character recognition                                                                                   & Python \& C++                      &      $\bullet$                           &            $\bullet$                     &                                 &                                 \\ 
{\sf linpack}                                              & \begin{tabular}[c]{@{}l@{}}Solves linear equations \\ using matrices\end{tabular}                               & Python                             &     $\bullet$                            &                                 &       $\bullet$                          &                                 \\ 
{\sf s3}                                                   & \begin{tabular}[c]{@{}l@{}}Download and then uploads \\ an object from one S3 bucket \\ to another\end{tabular} & Python                             &                                 &                                 &                                 &  $\bullet$                               \\ \bottomrule
\end{tabular}
}
\caption{Benchmark serverless functions. Each benchmark has certain resources that are more important than others, either for performance or to avoid function failure. Resources: C: CPU, P: parallelism, M: memory, N: network.}
\label{tab:benchmarks}
\end{table}

\subsection{Cost model}
\label{sec:cost-calculation}
The current pricing model of AWS Lambda charges users primarily based on GB per ms consumed by a function execution. Users only select the amount of memory allocated to each function instance. Thus, the precise per-vCPU and per-GB memory pricing necessary for execution cost calculation for this work are not available. Therefore, we adopt a simple cost model that assumes that CPU and memory are the two components contributing to the total cost of an instance.

We use the AWS instance pricing to calculate the per-vCPU and per-GB memory costs for different instance types. These costs are calculated by solving a system of linear equations for instances with the same CPU architecture type under the assumption that those instances would have the same per-GB pricing.
Each equation is of the form below that defines the instance price $P_{instance}$ (given as input) as a sum of per-vCPU cost $X$ and per-GB of memory cost $Y$:
\begin{equation}
    \alpha X_{vCPU} + \beta Y_{mem} = P_{instance}
\label{eq:cost_model}
\end{equation}

To create a well-defined system of equations, we use publicly available information from AWS to determine the number of vCPUs $\alpha$ and the amount of memory in GB $\beta$ for a sufficient number of equations.
For example, x86 instances m5, c5, and r5 are assumed to have the same per-GB memory cost $Y_{mem}$. Moreover, m5 and r5 instances have same CPU type. Thus, we have $X_{vCPU}^1$ for c5 and $X_{vCPU}^2$ for m5 and r5.
This yields a system of equations with 3 unknowns and 3 equations, using
$\alpha=2$ and $\beta=4,8,16$ for c5, m5, and r5 instances, respectively.
While we do not use r5 instances, its pricing information is used to solve the system.

We use the same approach to calculate per-vCPU and per-GB costs for ARM-based and AMD-based instances. The values for $\alpha$ and $\beta$ are the same for these instance families.

\section{How beneficial is flexible resource allocation?}
\label{sec:opportunity}
To understand the opportunities that current serverless offerings are missing, we start by characterizing the best execution time and cost for different resource allocation options. 

\subsection{Larger search spaces yield advantages...}

\begin{figure}
    \centering
    \includegraphics[width=0.9\linewidth]{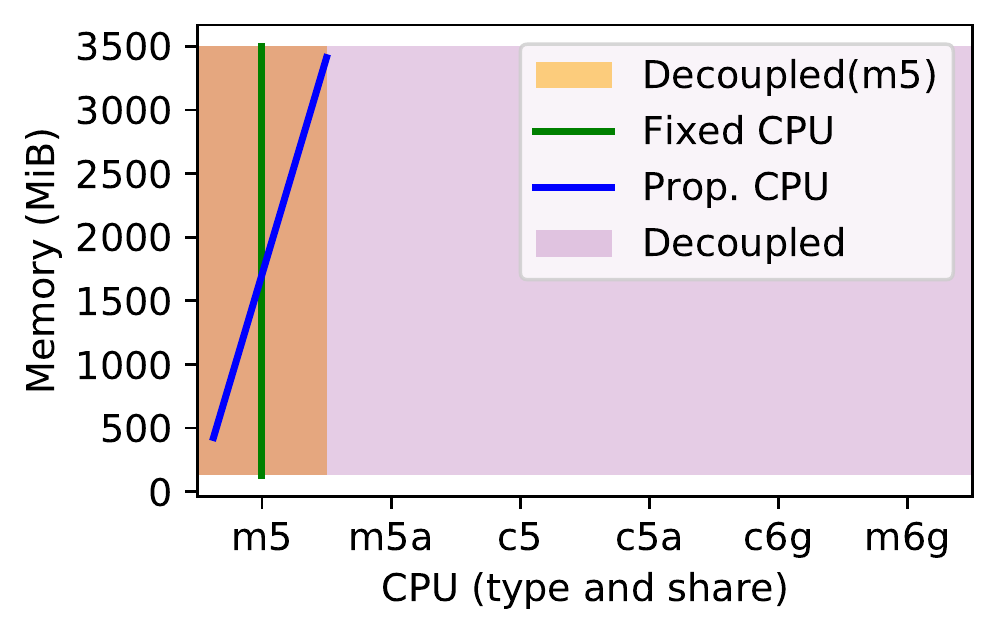}
\caption{Four strategies for configuring serverless functions. Each strategy defines a search space, corresponding to a subset of the possible resource allocation configurations.}
\label{fig:search-space}
\end{figure}

\smartparagraph{Setup.}
We consider four strategies for resource allocation, with an increasing level of flexibility. Each strategy corresponds to a subset of the configuration search space, as depicted in Figure~\ref{fig:search-space}.
The first three strategies assume a fixed instance type, which is the m5 type in our experiments.\\
\textit{Fixed CPU} allocates a single-vCPU for each function instance, whereas the memory is charged based on the average actual consumption. This strategy is inspired by Azure Functions.\\
\textit{Prop. CPU} allocates a share of CPU proportional to the amount of memory selected. This strategy is inspired by AWS Lambda and Google Cloud Functions.\\
\textit{Decoupled (m5)} decouples CPU and memory allocations. In this case, the search space includes all CPU and memory values in Table~\ref{tab:search-space}, but it uses only the default m5 type.\\
\textit{Decoupled} has the largest search space, encompassing all other strategies and covers the search space in Table~\ref{tab:search-space}. 

\begin{figure}
    \centering
    \includegraphics[width=\linewidth]{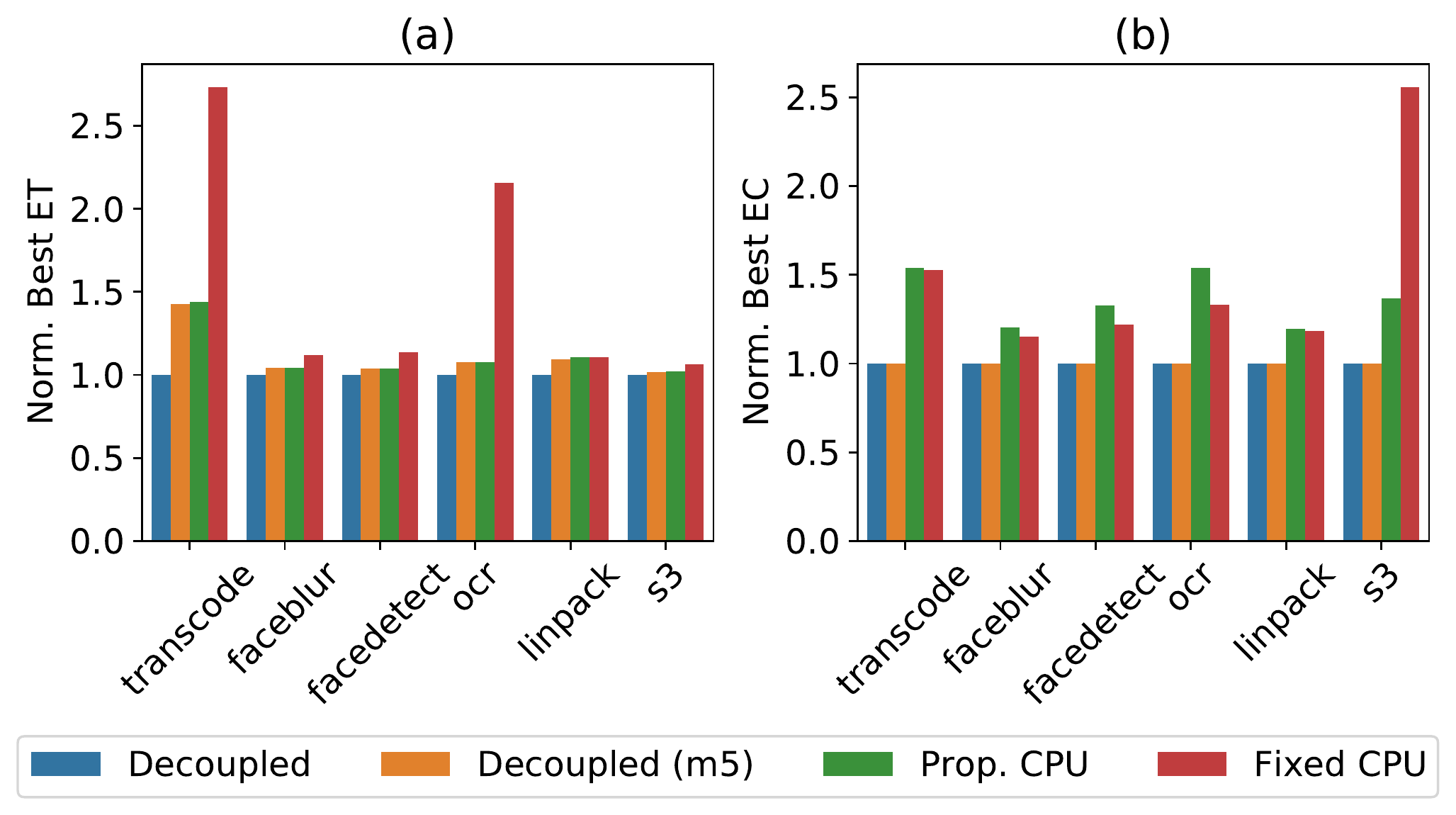}
\caption{Potential gains within each search space. The graphs show the best (a) Execution Time (ET) and (b) Execution Cost (EC) of each function across different search spaces, normalized to the overall best configuration.}
\label{fig:motivation-default-decoupled}
\end{figure}

\smartparagraph{Results.}
We measure the best execution time and execution cost for each benchmark function in each of the different resource allocation search spaces.
Figure~\ref{fig:motivation-default-decoupled} shows best execution time (ET) and execution cost (EC) of each strategy's search space normalized w.r.t. the best possible ones in \textit{Decoupled}, since its search space includes all others.

\smartparagraph{Observations.}
Figure~\ref{fig:motivation-default-decoupled}a shows that by using different instance types, \textit{Decoupled} can provide 5\%-40\% better execution time than \textit{Decoupled (m5)} and \textit{Prop. CPU}. In addition,  decoupling memory and CPU in \textit{Decoupled (m5)} is sufficient to provide 10\%-50\% better execution cost compared to \textit{Prop. CPU}, as shown in Figure~\ref{fig:motivation-default-decoupled}b.

\textit{Fixed CPU} leads to 2.7$\times$ and 2.1$\times$ higher execution time for {\sf transcode} and {\sf ocr}, respectively and 2.6$\times$ higher execution cost for {\sf s3}. This is because, for {\sf transcode} and {\sf ocr}, having 1 fixed vCPU per function invocation does not exploit available parallelism opportunities. For {\sf s3}, \textit{Fixed CPU} has higher execution cost because the function is not compute intensive and its execution time already plateaus with CPU share $<1$.

\smartparagraph{Takeaways:}
Using different instance types allows for improving the execution time, while decoupling CPU and memory enables potential improvements in execution cost compared to currently deployed resource allocation strategies.

\subsection{... and potential to reduce idleness}

\begin{table*}[]
\resizebox{0.98\textwidth}{!}{%
\begin{tabular}{l*{15}{>{\centering\arraybackslash}p{0.035\textwidth}}}
\toprule
\multicolumn{1}{c}{}                            & \multicolumn{3}{c}{Execution Time}                          & \multicolumn{3}{c}{$W_{t}$ = 0.25 \& $W_{c}$ = 0.75} & \multicolumn{3}{c}{$W_{t}$ = 0.5 \& $W_{c}$ = 0.5} & \multicolumn{3}{c}{$W_{t}$ = 0.75 \& $W_{c}$ = 0.25} & \multicolumn{3}{c}{Execution Cost}                          \\ \cmidrule(lr){2-4} \cmidrule(lr){5-7} \cmidrule(lr){8-10} \cmidrule(lr){11-13} \cmidrule(lr){14-16}
\multicolumn{1}{c}{}                            & \multicolumn{3}{c}{Threshold ($\theta$)}                    & \multicolumn{3}{c}{Threshold ($\theta$)}   & \multicolumn{3}{c}{Threshold ($\theta$)} & \multicolumn{3}{c}{Threshold ($\theta$)}   & \multicolumn{3}{c}{Threshold ($\theta$)}                    \\ \cmidrule(lr){2-4} \cmidrule(lr){5-7} \cmidrule(lr){8-10} \cmidrule(lr){11-13} \cmidrule(lr){14-16}
\multicolumn{1}{c}{\multirow{-3}{*}{Benchmark}} & 5\%                       & 10\%                      & 20\% & 5\%          & 10\%          & 20\%         & 5\%         & 10\%         & 20\%         & 5\%          & 10\%          & 20\%         & 5\%                       & 10\%                      & 20\% \\ \midrule 
{\sf ocr}                                        & 2        & 4        & \cellcolor[HTML]{34CDF9}5        & 1            & 1             & 2            & 1           & 1            & 4            & 1            & 3             & 4            & 1                         & 1                         & 2    \\ 
{\sf transcode}                                         & \cellcolor[HTML]{FD6864}0 & \cellcolor[HTML]{FD6864}0 & 2    & 2            & 2             & 2            & 2           & 2            & 2            & \cellcolor[HTML]{FD6864}0            & 2             & 2            & 1                         & 2                         & 2    \\ 
{\sf faceblur}                                         & 1                         & 2                         & 2    & \cellcolor[HTML]{FD6864}0            & 1             & 3            & 1           & 1            & 3            & 1            & 1             & 2            & \cellcolor[HTML]{FD6864}0                         & 3                         & 4    \\ 
{\sf facedetect}                                     & 1                         & 4                         & 4    & 1            & 1             & 3            & 2           & 3            & 3            & 2            & 3             & 3            & 1                         & 1                         & 3    \\ 
{\sf linpack}                                           & 2                         & 2                         & 4    & \cellcolor[HTML]{FD6864}0            & 2             & 3            & \cellcolor[HTML]{FD6864}0           & 2            & 3            & 1            & 1             & 4            & 1                         & 1                         & 3    \\ 
{\sf s3}                                                & 3                         & \cellcolor[HTML]{34CDF9}5 & \cellcolor[HTML]{34CDF9}5    & \cellcolor[HTML]{FD6864}0            & 1             & \cellcolor[HTML]{34CDF9}5            & 1           & 4            & \cellcolor[HTML]{34CDF9}5            & \cellcolor[HTML]{FD6864}0            & 4             & \cellcolor[HTML]{34CDF9}5            & \cellcolor[HTML]{FD6864}0 & \cellcolor[HTML]{FD6864}0 & 3    \\ \bottomrule

\end{tabular}
}
\caption{Number of alternative instance types with one or more resource allocation configurations with the performance metric within threshold ($\theta$) of the best configuration in the \textit{Decoupled} search space. The cells in red indicate cases where there is no alternative instance type able to reach within $\theta$\% of the best configuration. The cells in blue denote cases where all instance types have at least one configuration that provides performance within $\theta$\% of the best configuration.}
\label{tab:instance-type-performance}
\end{table*}

\smartparagraph{Setup.}
We now turn to another potential advantage of flexible resource allocation:
there may be cloud resources that are idle, but of the ``wrong'' instance type for a given function.
\textit{Decoupled} affords us the opportunity to use different instance types when doing so provides sufficient performance (when compared to the best configuration) according to different objectives.
We quantify these opportunities below.

\smartparagraph{Results.}
Table~\ref{tab:instance-type-performance} shows the number of instance types that have at least one configuration that is within $\theta$\% of the best \textit{Decoupled} configuration.
This table demonstrates the potential of using alternate instance types while providing comparable performance for different performance objectives for each benchmark function. The objectives are execution time (left), execution cost (right), and three weighted combinations of the two (denoted with $W_{t}$ and $W_{c}$ for execution time and cost, respectively).

\smartparagraph{Observations.}
We highlight two types of special cases in the table. The cells in red indicate cases where there is no alternate instance type able to reach within $\theta$\% of the best configuration. In other words, choosing a different instance type degrades performance by more than $\theta$\%.
The cells in blue denote cases where all instance types have at least one configuration that provides performance within $\theta$\% of the best configuration. 

The likelihood of being able to use idle resources clearly increases with a higher count of alternate instance types and depends on the function as well as the parameter $\theta$.

\smartparagraph{Takeaways: }
We found that except for two scenarios, there are opportunities to use idle instances of different types while providing performance within 10\% of the best configuration.

\section{Is automatically discovering good configurations possible?}
\label{sec:auto-opt}
As seen, more flexibility in the choice of resource allocation is beneficial.
However, in the presence of a large search space, the task of determining the right configuration can be daunting, requiring performance modeling and/or profiling, and particularly, it contrasts with the overarching goal of serverless --- to be a hassle-free computing service.
Thus, we now turn to explore the effectiveness of using \emph{black-box} optimization algorithms -- briefly reviewed below -- to automatically determine the right allocation of resources. 

\subsection{Background on optimization techniques}
\label{sec:bbo-algos}

Several recent works developed techniques for automatic cloud configuration~\cite{alipourfard17cherrypick, hsu2018scout, cloudconfigontrees2020, bilal2020vanir, venkataraman2016ernest}.
At the core of many of these approaches there lie various optimization techniques ranging from model-based optimization algorithms to sampling-based search techniques.
These approaches are also called black-box optimization algorithms because they consider the objective function as a black-box, which may be evaluated at specific points (using profiling runs) but there are no major assumptions that can be made about it.
In contrast, other works (e.g., Ernest~\cite{venkataraman2016ernest}) rely on analytical modeling to create a mathematical model that is dependent on the characteristics of the application and thus varies from one application to another. 

We believe that black-box optimization methods are a good fit for serverless functions, since the search space is large, function invocations can provide performance indicators, but the underlying objective function remains a black-box.
Even though the resulting configurations may be, in some cases, suboptimal compared to more precise analytical models, we believe that the advantage of being able to quickly reuse existing algorithms outweighs those limitations.

We next briefly review a range of optimization techniques that we use to automatically determine resource allocations for serverless functions.
Finally, we point out a relevant change when using these methods for the serverless scenario.

\smartparagraph{Model-based algorithms}
build a model of the underlying black-box objective function.
Following a comprehensive study~\cite{cloudconfigontrees2020} of black-box optimization algorithms, we adopt the best-performing method, which is Bayesian Optimization (BO).
We consider four variants of the surrogate model (required for approximating the objective function):
(1) Gaussian Processes (GP), (2) Gradient Boosted Regression Trees (GBRT),
(3) Random Forests (RF), and (4) Extra Trees (ET).
In all cases, we use the popular Expected Improvement (EI) as the acquisition function.

We use the Scikit-Optimize Python library~\cite{skopt} as a readily-available implementation of these algorithms and use its default parameters unless otherwise stated.
By default, based on previous findings~\cite{alipourfard17cherrypick}, we use three random initial samples to bootstrap model-based algorithms.

\smartparagraph{Sampling-based search techniques} sample the search space to find good configurations. These methods are simple to implement and easy to parallelize. We use both Random Sampling as well as Latin Hypercube Sampling (LHS)~\cite{mckaya1979lhs}. LHS samples the search space using a space filling design to generate near-random samples.
We use pyDOE~\cite{pydoe} to generate LHS samples.

\smartparagraph{Adapting to the serverless scenario.}
An issue that arises when using the above techniques out-of-the-box is that they can produce configurations on which the function fails because not enough memory is allocated.
At first, we attempted to address this by assigning a large value to represent the performance objective (e.g., execution time) of a failed function invocation.
However, that created a non-smooth underlying function, which affects the quality of the optimization.

Therefore, we instead deal with this issue by slicing the search space to remove from it the resource configurations with memory less than or equal to every memory configuration for which we determine that the function failed. This is based on the simple property that if a function fails for a certain memory limit, it is very likely to continue to fail with a lower memory limit. Thus, every time we record a function failure, the search space is dynamically reduced, removing all configurations that would lead to failure due to a lower memory limit. 

Next, we present the methods and results of our study, structured by its main findings.

\subsection{Optimization techniques are effective}
\label{sec:effectiveness}

\smartparagraph{Sampling-based vs. model-based algorithms.}
We first study how well black-box algorithms optimize the resource allocation as compared to the ground-truth best configuration in the \textit{Decoupled} search space.
We fix a budget of 20 trials for each method and report on the best execution time and cost found within those trials.
For the sampling-based methods (Random Sampling and LHS), this entails generating 20 samples. For the black-box optimization method (BO), we start with 3 initial samples and then use the acquisition function to repeatedly sample from the search space a series of configurations to test until the budget is matched. We repeat the optimization process 10 times using different random seeds (for sampling) and initial samples (for BO).

\begin{figure}
    \centering
    \includegraphics[width=\linewidth]{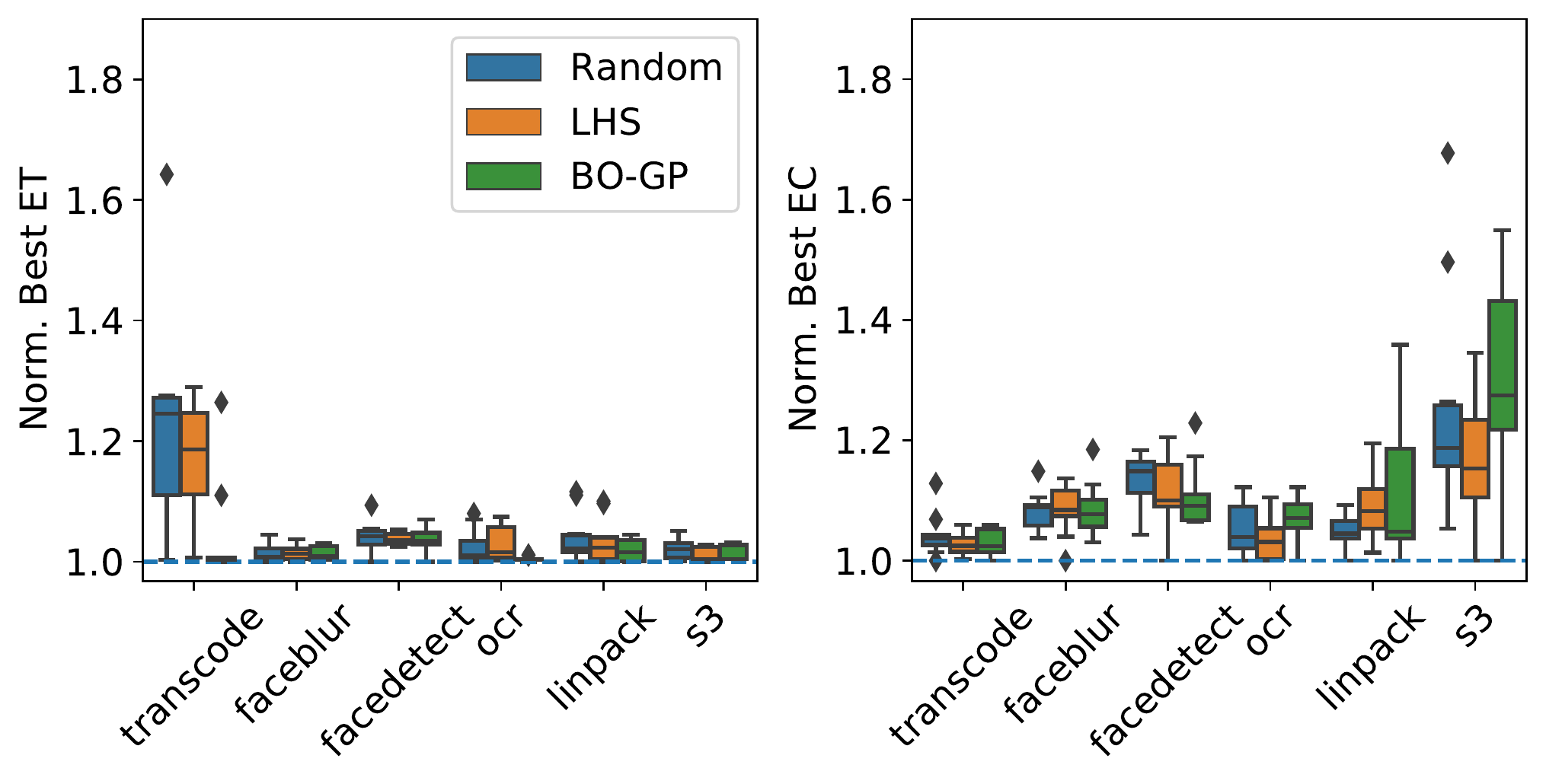}
    \caption{Performance of the best-found configurations for sampling-based and model-based algorithms.}
    \label{fig:sampling-or-modeling}
\end{figure}

Figure~\ref{fig:sampling-or-modeling} shows the best-found execution time and execution cost normalized w.r.t. the best configuration in the search space.
The boxplot captures the variation in the best-found configuration across different repetitions.
For these results, we only show the values for BO with GP, because this method outperforms other BO variants.

We observe that both sampling methods and BO with GP perform comparably in most cases, although GO with GP finds a better execution time for {\sf transcode}, whereas
sampling methods reach a better execution cost for {\sf s3}.
We obtained similar results for weighted combinations of execution time and execution cost (not shown due to space limitations).

We additionally note that, while sampling-based methods are simpler to use, model-based methods generate models that can provide predictions for yet-unseen configurations.
Therefore, given the overall good performance of model-based methods and the importance of predicting configurations with larger search spaces, we mainly use model-based methods in the rest of this paper.

\smartparagraph{Convergence speed of model-based algorithms.}
We now analyze how fast model-based methods converge towards the best configuration in the search space for each benchmark function. This analysis offers an early indication of how many optimization trials would be necessary as a baseline, as we later turn to online optimization in \S\ref{sec:online-perf}.

\begin{figure*}
    \centering
    \includegraphics[width=\textwidth]{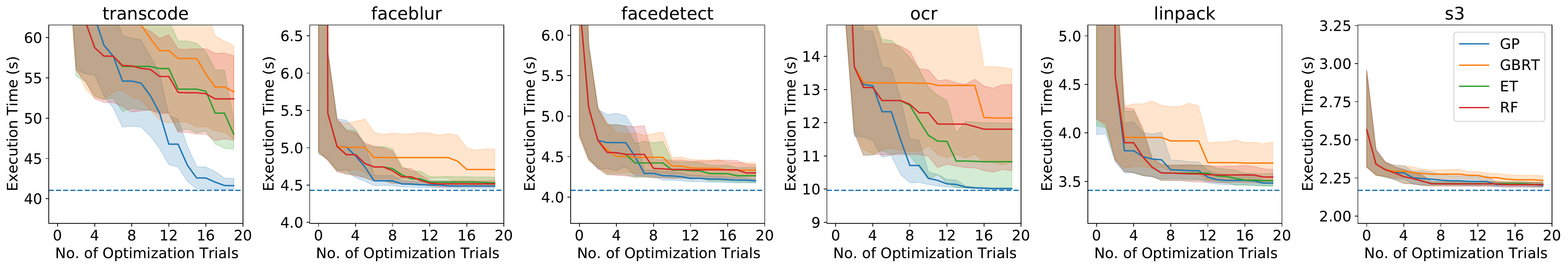}
    \caption{Execution time performance of the best found configuration as optimization by different BO variants progresses.}
    \label{fig:opt-et}
\end{figure*}

\begin{figure*}
    \centering
    \includegraphics[width=\textwidth]{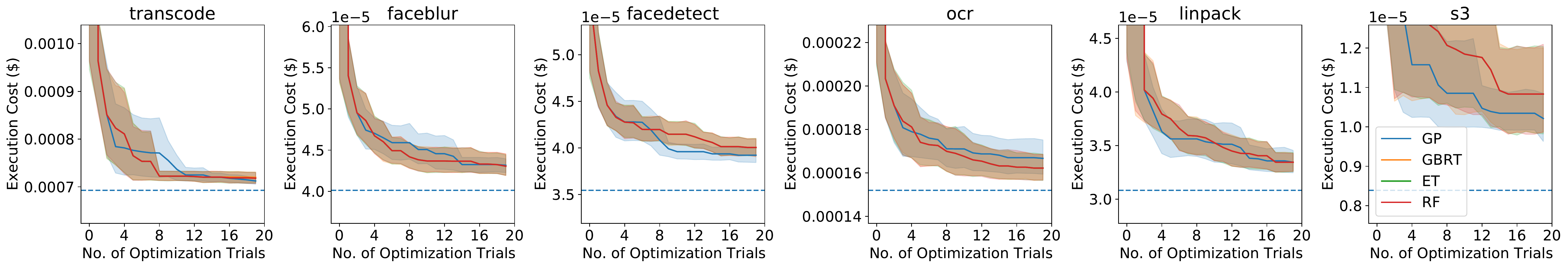}
    \caption{Execution cost performance of the best found configuration as optimization by different BO variants progresses.}
    \label{fig:opt-ec}
\end{figure*}

We run the four BO variants for 20 steps (including the 3 initial samples), using execution time and execution cost as the performance objective.
Figures~\ref{fig:opt-et} and \ref{fig:opt-ec} show the execution time and execution cost, respectively, of the best-found configuration as the optimization process progresses. 
The dashed lines denote the overall best execution time (resp.\ execution cost) in the search space. We repeat the experiment 10 times for each optimization method using different random initial samples. The shaded area is the $95^{th}$ percentile confidence interval.

With regard to execution time, while almost all optimization algorithms perform comparably, BO with GP overall tends to outperform other methods since it reaches configurations with comparable or better execution time compared to the second-best optimization method. In particular, BO with GP reaches within 5\% of the best execution time in 20 optimization trials, in all cases.

When optimizing for execution cost, the best configurations are harder to find and thus there is a larger gap between the best-found configurations and the overall best ones in the search space. This is particularly true for {\sf s3} and {\sf facedetect}. In 5 out of 6 benchmarks, BO with GP finds configurations that are within 20\% of the best execution cost in the search space. But for {\sf s3}, it only finds configurations that are within 30\%. However, similarly with optimizing for execution time, BO with GP either outperforms or performs comparably to other BO variants.

\smartparagraph{Takeaways: }
While both sampling-based search and model-based optimization methods find good resource allocation, we favor model-based methods since they can provide predictions for untested configurations as well. Different BO variants perform comparably when optimizing for execution time and execution cost, with an advantage for BO with GP. 

\subsection{Input data variation has modest influence}
\label{sec:input-dep}

\begin{figure*}
    \centering
    \includegraphics[width=\linewidth]{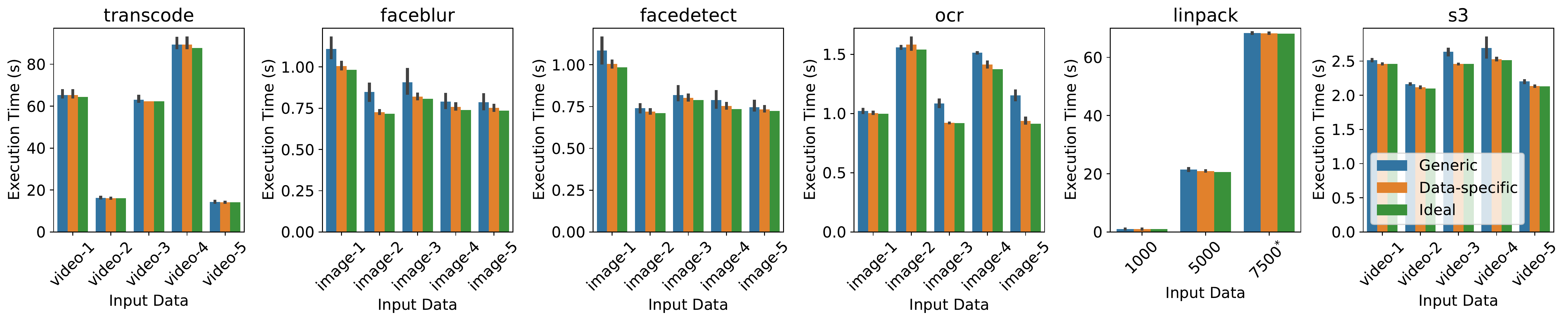}
    \caption{Execution time of the best configurations found by a generic optimization process (blue), a data-specific optimization process (orange), and the overall ideal configuration (green).}
    \label{fig:opt-across-data}
\end{figure*}

The performance of a function in general depends on its input data.
For instance, {\sf transcode} significantly depends on the input video dimension whereas {\sf facedetect} and {\sf faceblur} depend on the input image size.
Indeed, all of our benchmark functions have input data-dependent execution time.

Dealing with data dependence is challenging.
One approach would be to create data-specific performance models that account for the input data characteristics. However, this adds significant complexity to the optimization framework. First, the model needs to be sophisticated enough to encapsulate these performance-determining characteristics, which is especially difficult if the relation between input data and execution time is complex (e.g., if it has many modalities that require advanced profiling~\cite{rogora2020}). Second, even utilizing such a performance model is difficult because, at invocation time, the serverless framework would need to evaluate with minimal overhead the input data and route accordingly. 

A simpler approach instead is to use a generic optimization process to create a model using representative input data samples. This is based on the intuition that, even though the absolute performance may vary, a good configuration for one input data sample might also be a good configuration for other input data samples.

We therefore study to what extent the resource configuration depends on input data and the effectiveness of this second approach.
To this end, we use the methods in \S\ref{sec:effectiveness} so that, for each function, we use the default input to create a single generic model as well as 10 input-specific models, one for each input sample.

Figure~\ref{fig:opt-across-data} contrasts the performance of the best-found configuration for the two model types -- generic (blue) vs. input-specific (orange) -- for each input, together with the overall best configuration (green) in the search space (for that input). For clarity, we only show the results of 5 out of 10 input samples (except for {\sf linpack}).
We show the optimization scenario for execution time and note that optimizing for execution cost obtains similar results.

In our experiments, using input-specific models provides up to 20\% better execution time. {\sf linpack} with a $N=7500$ matrix is a special case since it requires a large memory and the optimization process using a default input leads to lack-of-memory failure in 3 out of 10 repetitions of the optimization process (which we exclude).

This modest improvement in performance comes with significantly increased complexity. In addition, in a real setting, each input cannot have its own model. (We leave it to future work to improve on creating more sophisticated performance models.)
Thus, we conclude that our experiments expose a trade-off between framework complexity and potential gains (up to 20\%) due to accounting for input-specific characteristics. Consequently, in the rest of the paper, we adopt the generic models.

\smartparagraph{Takeaways: }
While the performance of a serverless function in general depends on its input data, our experiments indicate that configurations that are good for one input sample are also good for others. Input-specific optimization leads up to 20\% improvement in execution time, but because input-specific performance models are more complex to create and maintain, this trade-off needs to be considered carefully.

\subsection{Online optimization is feasible}
\label{sec:online-perf}
The optimization of resource configurations can occur in two ways: (1) offline or (2) online.
Offline optimization requires running the optimization process described in \S\ref{sec:effectiveness} upon function deployment with representative input samples.
Conversely, online optimization exploits function invocations in production as trials of the optimization process. 
However, in the online scenario, it is important to reduce the possibility of performance degradation due to trials with bad configurations.
Thus, we now analyze which optimization methods lead to fewer degraded runs during optimization. 

\begin{figure}
    \centering
    \includegraphics[width=\linewidth]{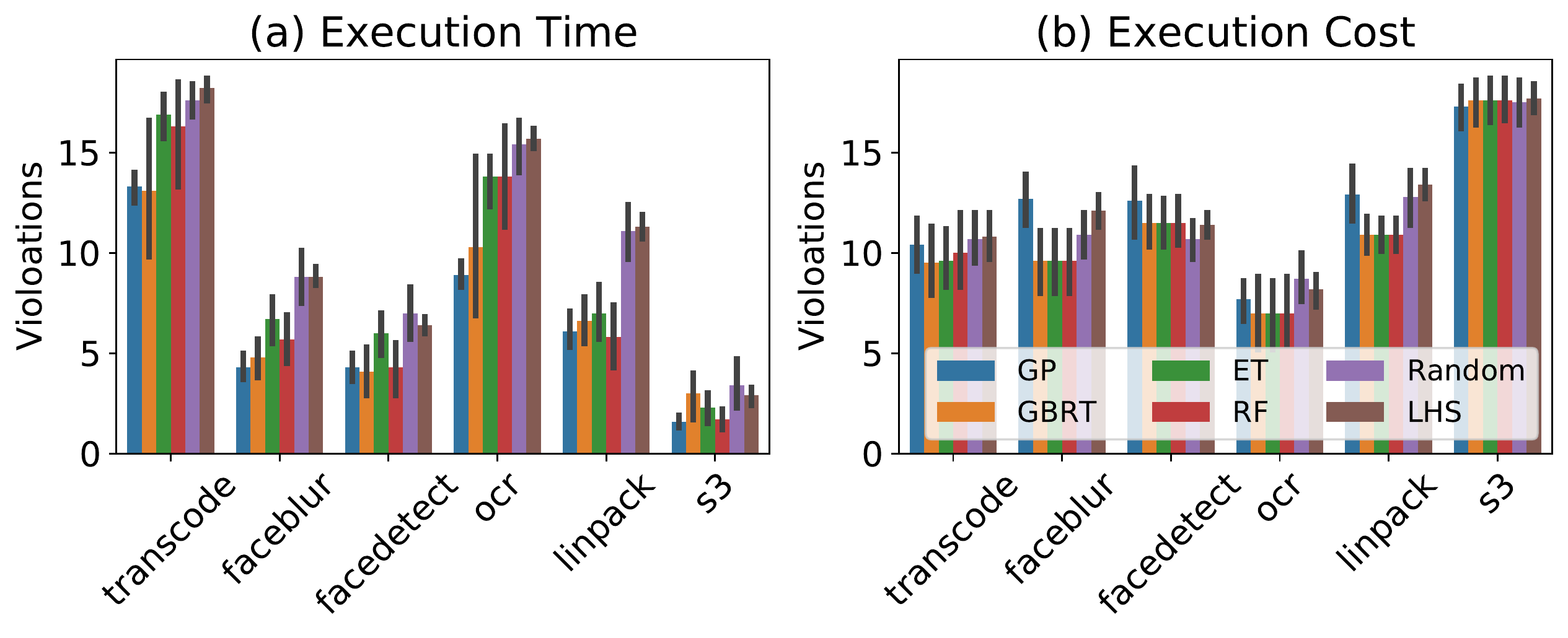}
    \caption{Average number violations during online optimization for (a) execution time (ET) and (b) execution cost (EC).}
    \label{fig:violations}
\end{figure}

Figure~\ref{fig:violations} shows the average number of violations during 10 repetitions of the optimization process.
Here we consider a violation when the performance objective is at or above 1.5$\times$ the objective value for the best configuration in the search space. On average, BO with GP has a number of violations comparable to sampling-based search techniques for execution cost, but it has a lower number of violations for execution time. Overall, sampling-based search techniques have a slightly higher number of violations compared to model-based methods. 

\smartparagraph{Takeaways:}
BO with GP has slightly lower average number of violations for execution time compared to other methods. For execution cost, it has slightly higher number of violations compared to other BO variants, but overall it leads to fewer violations than sampling-based search methods.

\subsection{Resource allocation models can predict performance of untested configurations}

It is one thing to converge to the best performing configuration in the search space, but predicting the performance of an untested configuration is another.
In this context, a low prediction accuracy would suggest that sampling-based search methods, which are simpler, may be sufficient after all.
Therefore, we now analyze how well model-based methods can predict the performance objective across configurations in two different scenarios:
(1) over the entire \textit{Decoupled} search space (except the failed runs), and (2) when the configuration prediction is restricted to match a particular instance type.
This second scenario is relevant in the context of helping the cloud provider utilize idle resources of different instance types while providing predictable performance (as we elaborate in \S\ref{sec:different-instance-types}).
We repeat every measurement 10 times and report average and $95^{th}$ percentile confidence interval of the error metric.

\begin{figure}
    \centering
    \includegraphics[width=\linewidth]{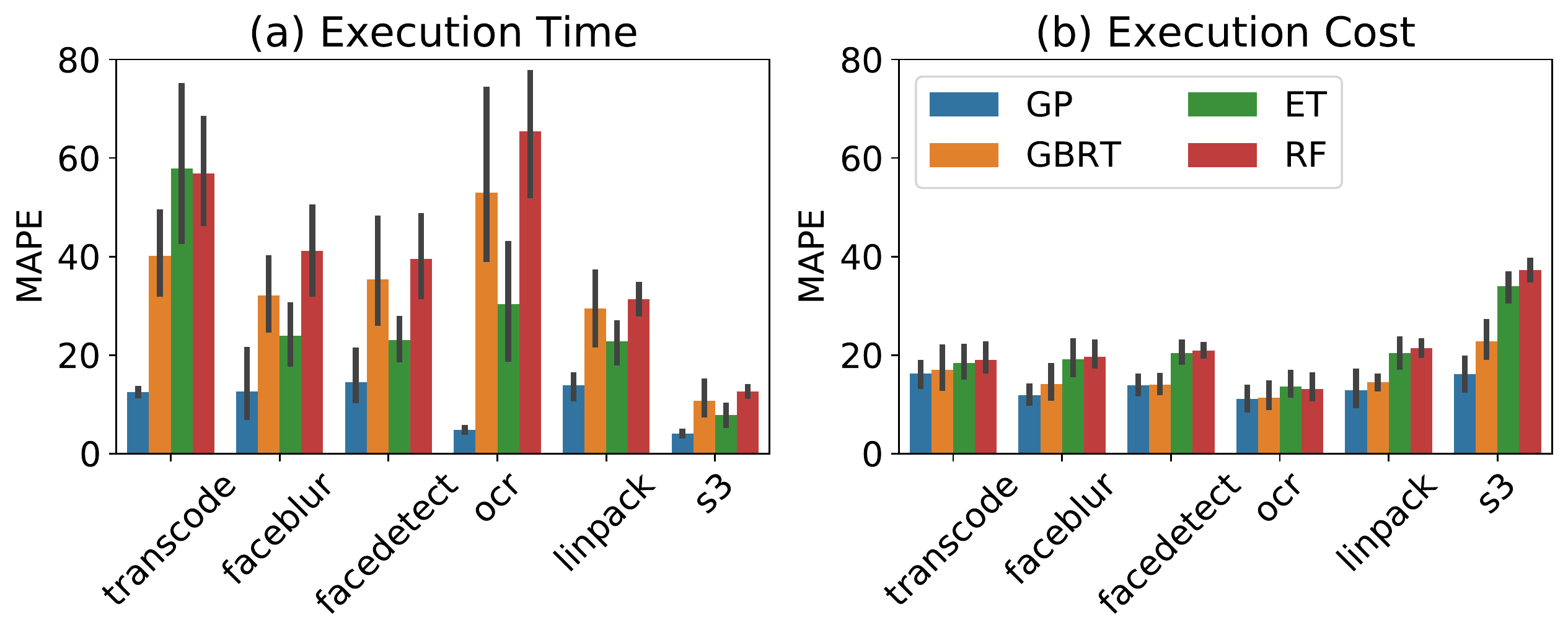}
    \caption{MAPE for different benchmarks and optimization methods when optimizing for (a) execution time and (b) execution cost. }
    \label{fig:errors}
\end{figure}

\smartparagraph{Scenario 1.}
Figure~\ref{fig:errors} shows, for each BO variant, the Mean Absolute Percentage Error (MAPE) across all configurations between the actual execution time/cost and the predicted value. We observe that BO with GP has lower error than other optimization algorithms. Compared to other variants, BO with GP, on average, has up to 16$\times$ and 2.3$\times$ lower MAPE for execution time and execution cost objective, respectively.

\begin{figure}
    \centering
    \includegraphics[width=\linewidth]{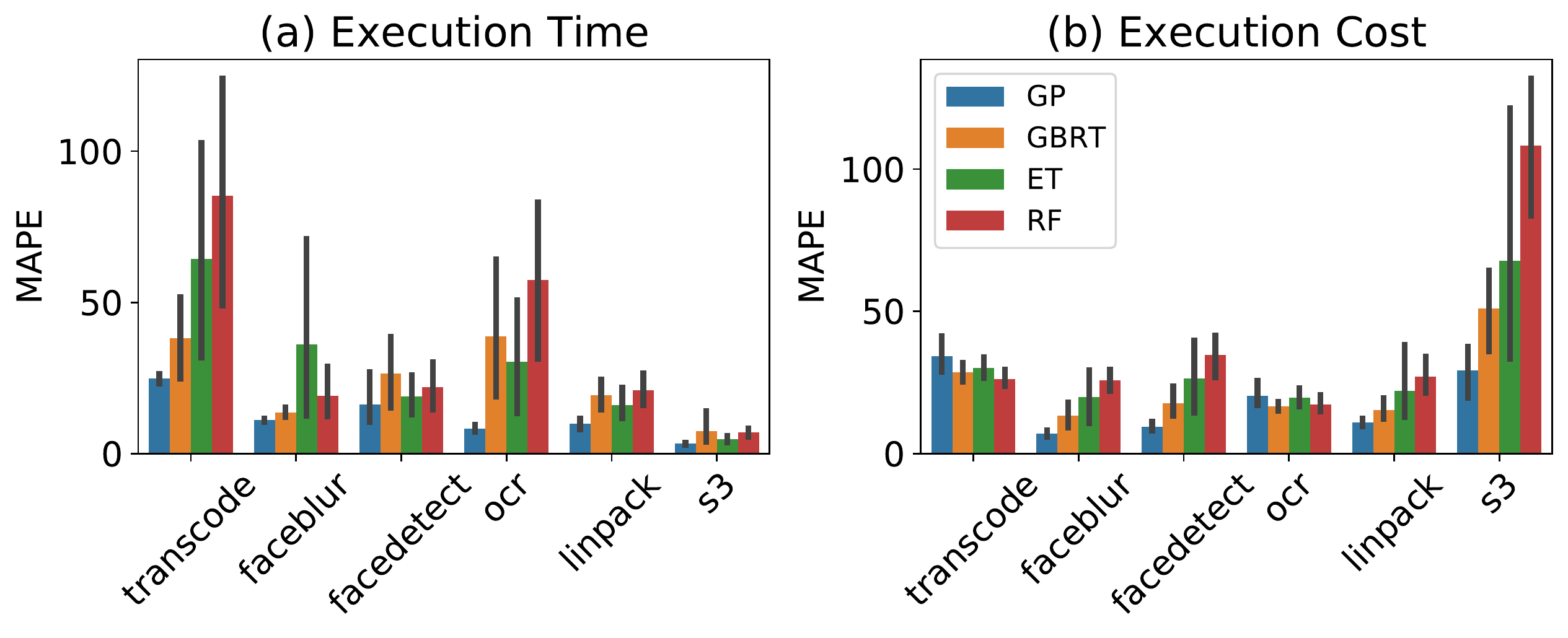}
    \caption{MAPE for different benchmarks and optimization methods when comparing the best configuration for each instance type, when optimizing for (a) execution time, and (b)  execution cost.}
    \label{fig:error-across-types}
\end{figure}

\smartparagraph{Scenario 2.}
Figure~\ref{fig:error-across-types} shows the MAPE across the best predicted configuration for each instance type. Similar to the previous case, BO with GP generally outperforms other BO variants in most cases, except for {\sf transcode} and {\sf ocr} when optimizing for execution cost. BO with GP, on average, has up to 7$\times$ and 3.5$\times$ lower MAPE than other variants for the execution time and execution cost objective, respectively. 

\smartparagraph{Takeaways: }
BO with GP not only works well when it comes to convergence towards the best configuration (as shown in \S\ref{sec:effectiveness}) but overall has lower error compared to models created using other BO variants. In our experiments, models built with BO with GP provide up to 16$\times$ lower MAPE than models build with other BO variants. This means that performance models built with BO with GP are better at predicting the performance of untested configurations. 

\section{Automatic resource allocation from the provider's perspective}

Having looked at the potential gains from flexible resource allocation, and the effectiveness with which we can exploit these gains using black-box optimization algorithms, in this section we discuss how the cloud provider can not only expose this to users without unduly complicating the `serverless' interface, but also leverage model predictions to opportunistically select available instance types that can reduce costs without significantly sacrificing performance.

\subsection{On the interface between user and provider}
\label{sec:mo-interface}
Given the performance and cost benefits of flexible resource allocations, 
the provider could expose the knobs for fine-grained resource configuration (selecting instance type, memory and CPU allocation separately). While possible, this would, however, shift the complexity of configuration
selection back to the user, and negate one of the big advantages of serverless: its simplicity.
A more sensible alternative would be for the provider to abstract away that interface and allocate resources automatically and transparently. 
Our automatic exploration of the configuration space
allows the best of both worlds: a simple, high-level interface that exposes to the user the cost/performance benefits of decoupled resource allocation.

We describe three ways to allow users to select a trade-off between execution time and cost: 1) Providing configurations from the predicted Pareto front, 2) Weighted multi-objective optimization~\cite{arora2004introtoopt} to provide best configurations for different weights, and 3) Hierarchical multi-objective optimization~\cite{arora2004introtoopt} to satisfy a user-provided trade-off constraint.

\begin{figure}
    \centering
    \includegraphics[width=0.7\linewidth]{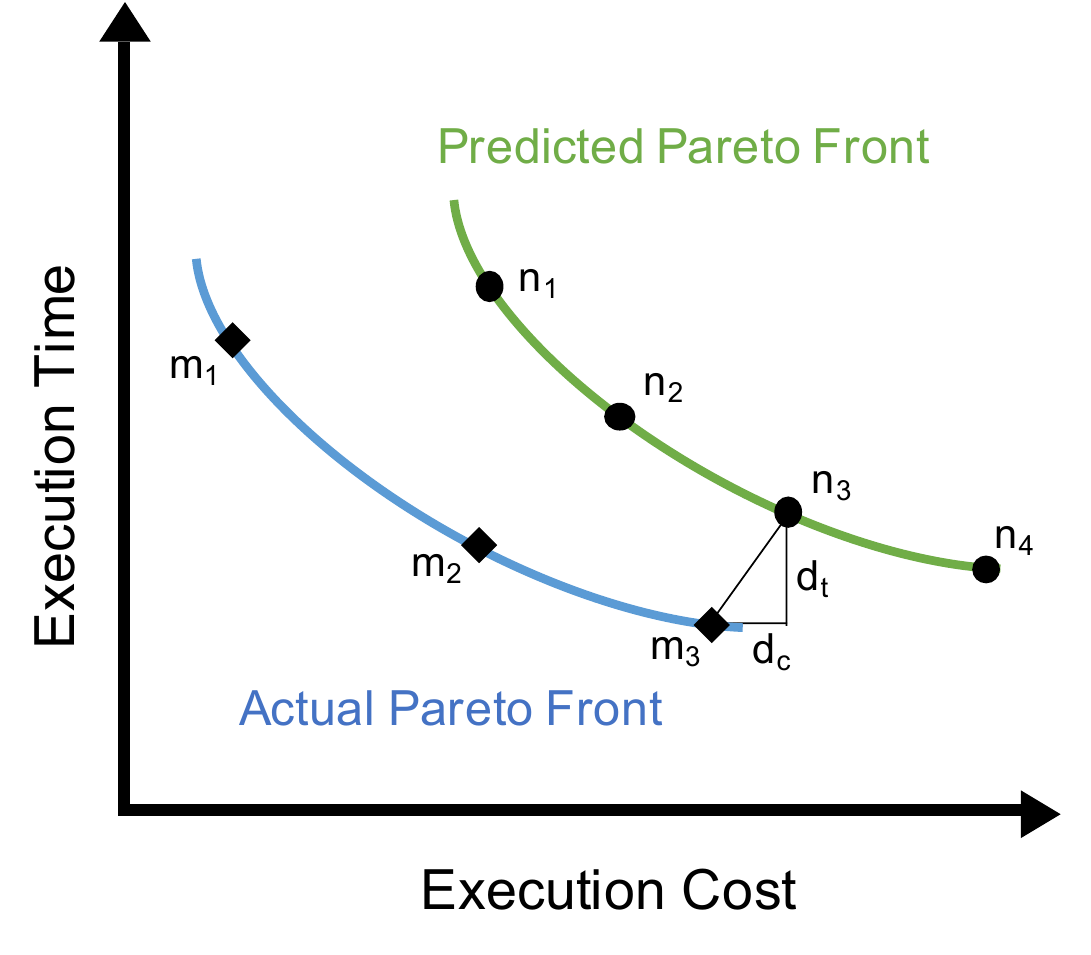}
    \caption{Method for measuring the distance between the configurations in the predicted and actual Pareto front. The distance is split into execution time and execution cost components and measured between each configuration in the predicted Pareto front and the nearest configuration in actual Pareto front.}
    \label{fig:pareto-distance-measure}
\end{figure}

\begin{figure}
    \centering
    \includegraphics[width=0.8\linewidth]{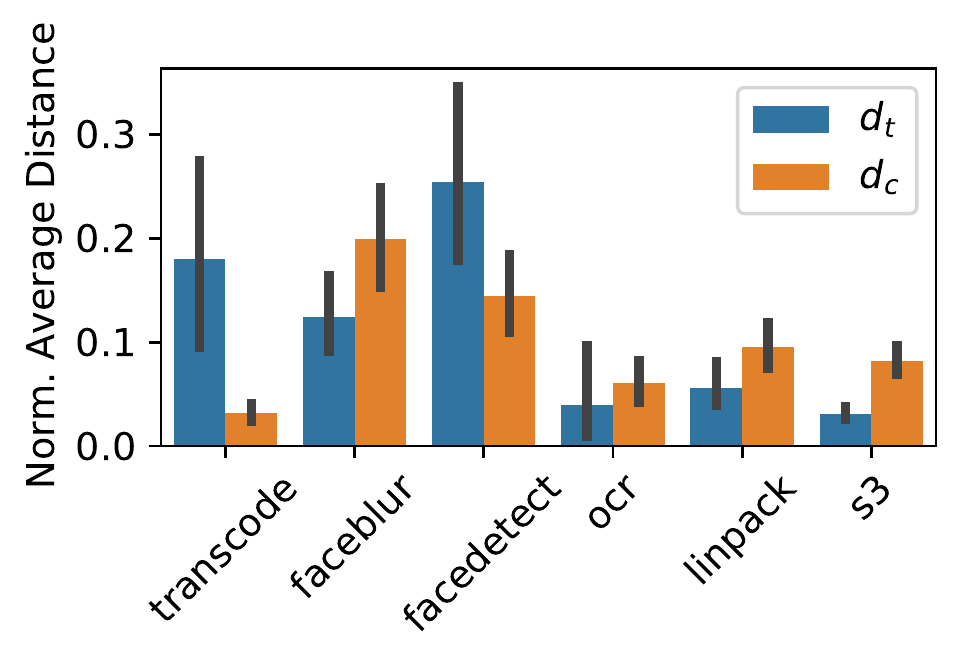}
    \caption{Normalized average distance between the points in the predicted Pareto front and the actual front for the default input. We show normalized execution cost ($d_c$) and execution time ($d_t)$ components of the distance separately.}
    \label{fig:pareto-distance}
\end{figure}

\smartparagraph{Pareto front: }
We can use the predictions from the black-box model to create a Pareto front and expose to users the configurations that have different trade-offs between execution time and execution cost. The predicted Pareto front is created by normalizing the execution time and execution cost so that the values of the two objectives are on a similar scale. Since we do not know the actual minimum value of the objectives, we use the minimum values observed while optimizing execution cost and execution time to perform normalization.
Therefore, two models have to be trained to create a Pareto front. 

To assess the effectiveness of using the configurations from the predicted Pareto front, we measure the distance of those configurations from the predicted Pareto front to the nearest configuration in the actual Pareto front, as shown in Figure~\ref{fig:pareto-distance-measure}. We measure the distances in terms of normalized execution time ($d_{t}$) and cost ($d_{c}$) separately. $d_{t}$ and $d_{c}$ for each configuration are normalized using the corresponding objective value for the nearest configuration in the actual Pareto front.

Figure~\ref{fig:pareto-distance} shows, for each function, the average distance between the points in the predicted Pareto front and the actual Pareto front for the default input.
Note that the prediction error in the model leads to $d_{t} > 0, d_{c} > 0$. In our experiments, the average difference between the configurations in predicted Pareto front and actual Pareto fronts is up to 20\% (cost) and 25\% (time). 

The interface to the user exposes the small set of configurations (between 2 and 10) in the Pareto front, with the corresponding predicted cost and execution times. 

\smartparagraph{Weighted multi-objective optimization:}
With weighted multi-objective optimization, the
cloud provider can select relative weights for execution time ($W_{t}$) and execution cost ($W_{c}$), where $W_{c}$ = $1-W_{t}$. Using these weights, the cloud provider can form a weighted objective function $F_{w}(X)$ for a configuration $X$, from normalized objective functions for execution time ($F_{t}(X)$) and cost ($F_{c}(X)$):
\begin{equation}
\label{eq:weighted-obj-function}
    F_{w}(X) = W_{t}\times \frac{F_{t}(X)}{B_{t}} + W_{c}\times \frac{F_{c}(X)}{B_{c}}
\end{equation}

\noindent where $B_{t}$ and $B_{c}$ are the minimum values for execution cost and execution time objectives, found during the optimization process for $F_{t}(X)$ and $F_{c}(X)$.
To simplify the process for the user, we pre-train three models with $W_{t} \in \{0.25, 0.5, 0.75\}$.
The two models trained first for execution time and execution cost translate to $W_{t}=1$ and $W_{t}=0$, giving a total of five models and five best configurations (one configuration suggested by each optimization process) for the user to choose from.

\begin{figure*}
    \centering
    \includegraphics[width=0.9\linewidth]{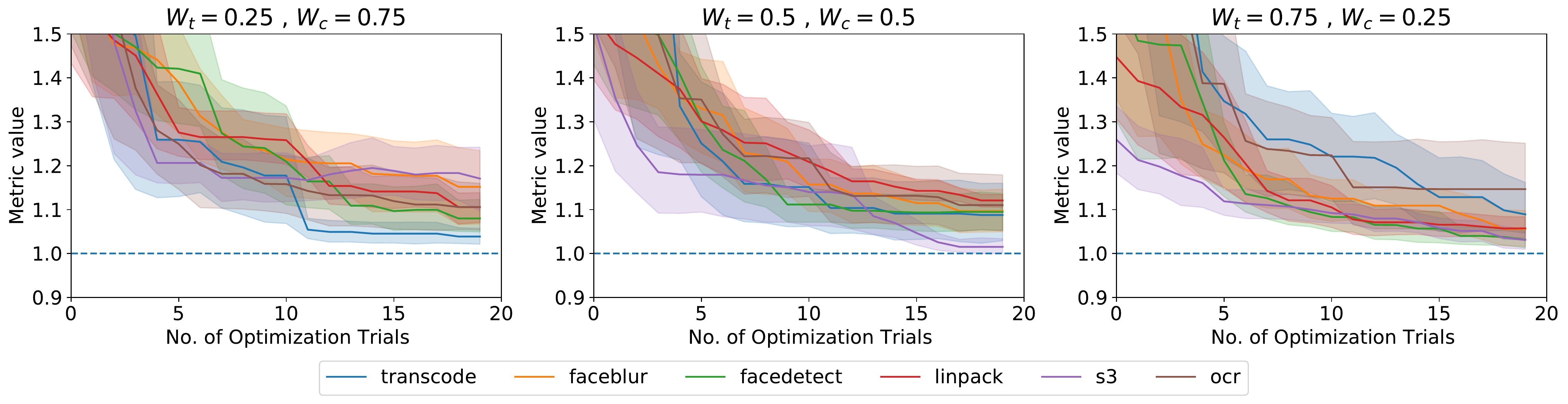}
    \caption{Convergence of the optimization process (BO with GP) for all the benchmarks and different weights for execution time and execution cost.}
    \label{fig:weighted-mo}
\end{figure*}

Figure~\ref{fig:weighted-mo} shows the best configurations found by the weighted multi-objective optimization (using BO with GP) as we perform more optimization trials. We can see that even in the weighted multi-objective setting, for most cases, the optimization process is able to find configurations that are within 20\% of the best configurations in the search space after 20 optimization trials. While not shown here, we increased the number of trials and observed that within 40 trials, BO with GP is able to find configurations with performance within 5-10\% of the best configuration, for all cases. 

The interface is very similar to the Pareto front one: in this case, the user can choose between at most 5 configurations, based solely on their predicted cost and performance.

\smartparagraph{Hierarchical multi-objective optimization: }
In hierarchical multi-objective optimization, we first optimize one of the objective functions (primary objective). Then, the optimized model can be used to find configurations that minimize the value for the second objective function (secondary objective) while degrading the primary objective value by at most a user-defined amount ($\theta$). This optimization process may be easier for users to reason about. Once they are shown the best value for the primary objective function, they can decide if they are willing to degrade that value by a certain percentage to improve the secondary objective function's value. For example, users can choose to increase the execution time by 20\% compared to the best execution time found as long as the corresponding execution cost decreases. 

\begin{figure}
    \centering
    \includegraphics[width=0.9\linewidth]{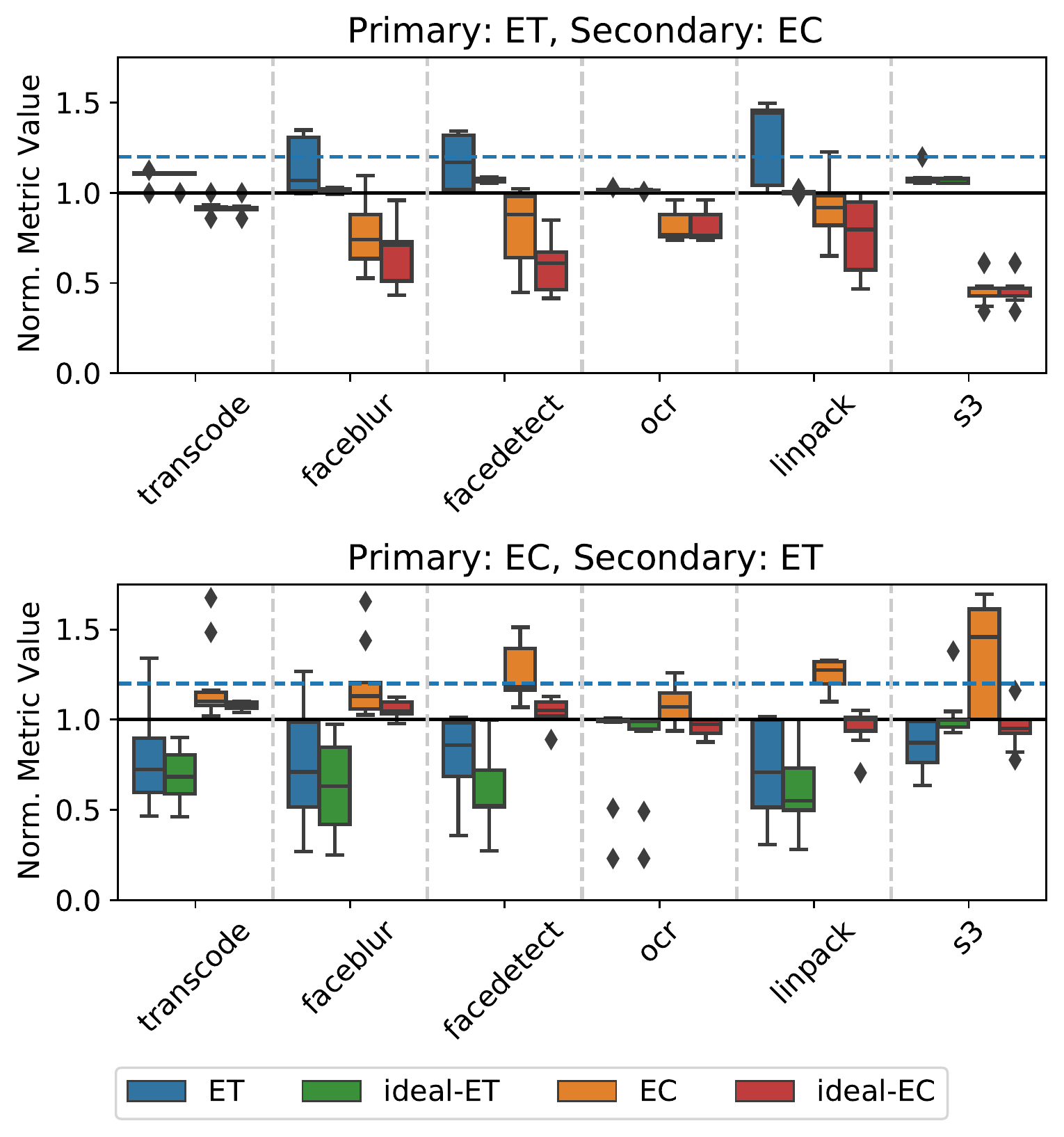}
    \caption{Normalized values for execution time and execution cost after a hierarchical optimization. ET/EC and ideal-ET/ideal-EC represent the best configuration using the prediction model and with oracle-like knowledge, respectively.}
    \label{fig:hierarc-mo}
\end{figure}

Figure~\ref{fig:hierarc-mo} shows the normalized value for execution time and execution cost metrics after the hierarchical optimization (satisfying user's constraints) for the two combinations of primary and secondary objective functions. We used a threshold of $\theta=20\%$ for this experiment. ET/EC and ideal-ET/ideal-EC represent the best configuration using the prediction model and oracle-like knowledge. The normalization is done w.r.t.\ the best configuration found after optimizing the primary objective only. The dashed line shows the user-specified degradation threshold for the primary objective. In some cases, we can see that prediction error leads to a higher degradation of the primary objective than the threshold. But in other cases, hierarchical optimization using the performance models performs comparably to the ideal case. 
Unlike weighted multi-objective optimization, only one model is trained for hierarchical multi-objective optimization.

\smartparagraph{Takeaways: }
While not the final answer, these three options give the user access to a much broader space of configurations than current offerings, without requiring the user to deal with complex resource allocation choices. In fact, they shift the language from \emph{resources} to 
\emph{outcomes}: performance and cost. They also represent different trade-offs in terms
of simplicity and effectiveness.
With Pareto front and weighted multi-objective optimization users would simply get a small set of cost-performance tuples that they can select from.
In our experiments, both methods found configurations which were up to 30\% worse than
the best in both cost and performance. The hierarchical optimization offers a more explicit prioritization, but users have to choose the threshold themselves. It also has good results: for an increase of roughly 20\% in the primary metric, a reduction of up to 50\% in the other. They also present different costs for the provider. For Pareto front and hierarchical multi-objective optimization we need to train 2 and 1 models, respectively. For weighted multi-objective optimization, the number of models we need to train depends on the number of weighted combinations of ET and EC. A full evaluation of these interfaces would require user studies and a more thorough cost analysis, and we leave it for future work.

\subsection{Cost-performance trade-off while using different instance types}
\label{sec:different-instance-types}
Table~\ref{tab:instance-type-performance} shows that there is potential for using different instance types while providing performance within a certain margin of the best found configuration. We now evaluate how effectively we can utilize that potential. In particular, we measure this benefit by translating utilization of idle resources into a cost decrease. Similarly to spot instances, we assume that a serverless instance type with many idle instances is assigned a lower cost, to incentivize the utilization of the idle resources. We assume that the spot pricing for the serverless instance will decrease the per-CPU and per-GB cost to a fraction of the original price. 

\begin{figure}
    \centering
    \includegraphics[width=0.8\linewidth]{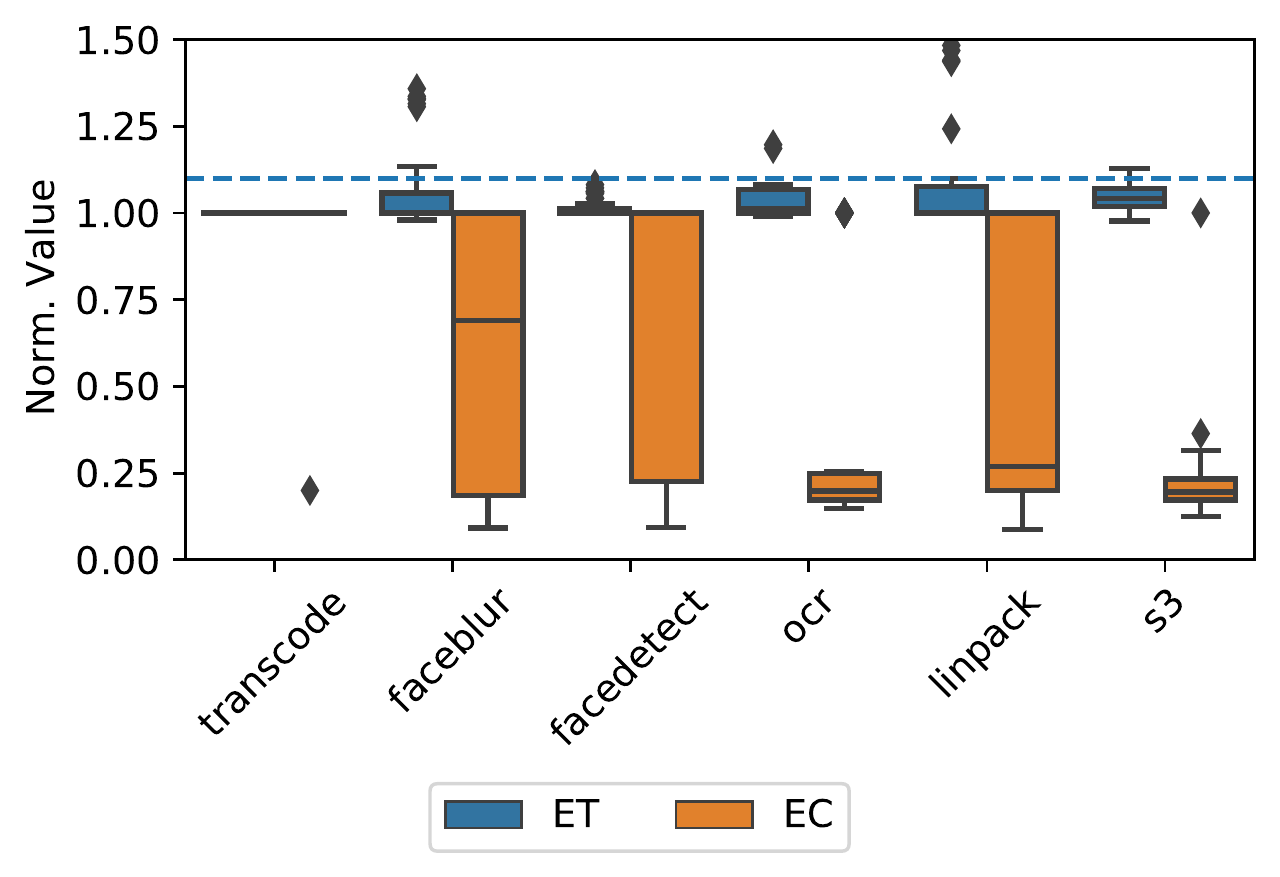}
    \caption{Reduction in cloud provider costs while keeping the objective function value within 10\% of the best found configuration in the search space. Cost reduction is based on model predictions for the best configurations of each instance type, assuming an 80\% discount in pricing for idle resources.}
    \label{fig:cost-benefit}
\end{figure}

Figure~\ref{fig:cost-benefit} shows the decrease in deployment cost that the cloud provider can observe by utilizing the best configurations for each instance type  predicted by the model. For this figure, we assume that spot pricing is 20\% of the normal pricing. Figure~\ref{fig:cost-benefit} shows the decrease in execution cost while the performance model is predicting configurations that are within 10\% of the execution time (marked by the dashed line) of the best found configuration. The figure shows the normalized value for execution time and execution cost w.r.t.\ to the best configuration found by the optimization process.

We can see from Figure~\ref{fig:cost-benefit} that, by using the predicted best configurations of other instance types, we can achieve between 25-75\% reduction in execution cost, on average, for different benchmarks. This execution cost reduction comes at  $<10\%$ increase in execution time, on average. There are outliers where the execution time penalty is up to 50\% because of prediction error. The main exception is the {\sf transcode}, for which there are very few execution cost reduction options available, as shown in Table~\ref{tab:instance-type-performance}.

\smartparagraph{Takeaways: }
In our experiments, we found that some configurations utilize different instance types but provide performance similar to the best configuration in the search space. A cloud provider can exploit this behavior and use prediction models to achieve lower execution costs (by using idle resources) while providing comparable execution time to the best found configuration. Even with prediction error, we show that it is possible to significantly reduce costs while delivering performance within 10\% (on average) of the execution time of the best found configuration.

\section{Design space}

\begin{figure*}
    \centering
    \includegraphics[width=\linewidth]{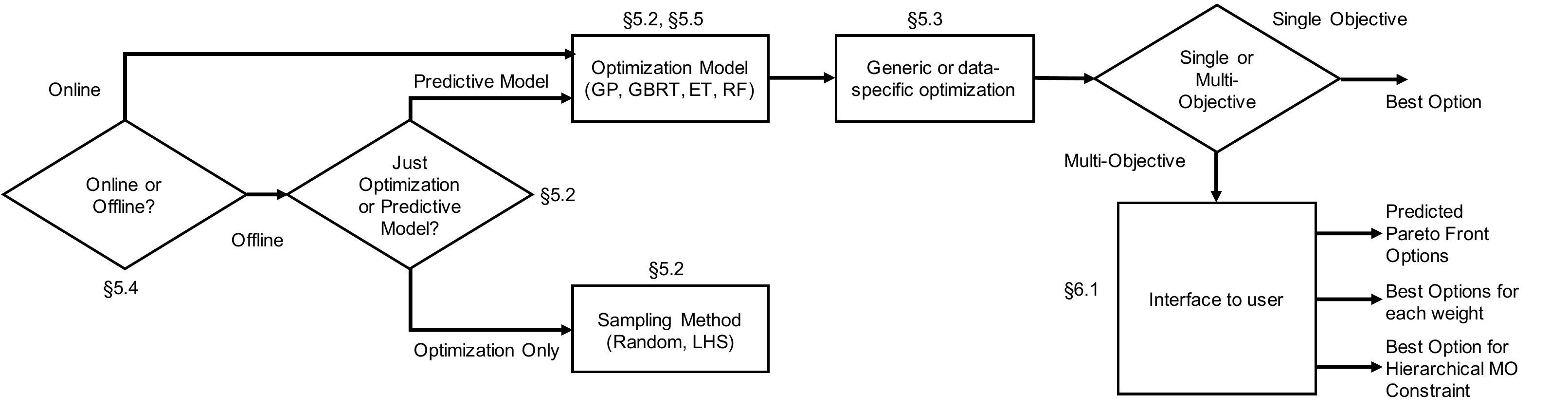}
    \caption{Systematic set of design choices of a system providing automatic resource allocation for serverless functions.}
    \label{fig:design-space}
\end{figure*}

Figure~\ref{fig:design-space} shows the design choices that should be considered when developing an automatic resource allocation system for serverless functions. First, the designer has to decide between providing offline or online optimization. If offline optimization is desired, then both search-based (sampling) and model-based optimization are valid choices. But with online optimization, we recommend a model-based approach. We found that Bayesian Optimization with Gaussian Processes performs better than other optimization algorithms that we tested, as it converges to the best configuration faster and has lower prediction errors than other BO variants.  

After deciding on the optimization algorithm, a designer would need to decide whether to create a data-specific or a generic optimization model. A data-specific optimization process might provide better performance but with added complexity. Irrespectively of whether the model is data-specific or not, one has to decide whether to provide single objective or multi-objective optimization options. A single objective optimization would provide the user with the option of either finding the configuration with the lowest execution time or cost. For multi-objective optimization, in turn, we discussed three options: 1) Pareto front, 2) weighted multi-objective optimization, and 3) hierarchical multi-objective optimization. Hierarchical optimization is the most intuitive for the end-user to reason about, since it allows the user to select a given trade-off the user is willing to accept.

\section{Related work}
\smartparagraph{Serverless computing.}
Several recent papers demonstrated the advantages of serverless computing by applying this paradigm to a series of applications, ranging from data analytics~\cite{muller2020lambada, pu2019shuffling}, DAG processing~\cite{carver2020wukong},  video transcoding~\cite{fouladi2017encoding}, compilation~\cite{fouladi2019laptop}, machine learning~\cite{carreira2019cirrusml, feng2018serverlessdnn, wang2019serverlessml} and more~\cite{ao2018sprocket, singhvi2020snf}.
Within this general area, there are a few recent proposals that aim to optimize the cost of serverless functions~\cite{elgamal2018costless, sedefouglu2021costopt, spillner2020functionmemory, eismann2020sizeless}. Sizeless~\cite{eismann2020sizeless} and~\cite{sedefouglu2021costopt} use a regression model to minimize execution cost of AWS lambda functions.~\cite{spillner2020functionmemory} uses memory tracing information to collect memory utilization metrics to find the right memory allocation for functions. Costless~\cite{elgamal2018costless} uses function fusion, splitting the function between edge and cloud, and allocating memory resources for a sequence of functions to optimize cost. However, these works are limited to the resource allocation strategy that AWS Lambda or other existing serverless offerings expose at the moment. In contrast, we take a step back and analyze the broad space of possible fine-grained configurations, and rethink the interface of services like AWS Lambda. 

Several works gain insights into the characteristics of public serverless offerings by creating experiments and benchmarks to test the behavior of these platforms~\cite{wang2018peeking,wen2020understanding,maissen2020faasdom, yu2020serverlessbench}. However, they do not analyze the space of possible configurations beyond the current offerings and their effects. 

HarvestVMs~\cite{ambati2020harvestvm} create flexible VMs that can grow and shrink based on the available unallocated resources in an underlying server, allowing cloud providers to utilize their resources more efficiently. Our work is complementary since it can be used in conjunction with HarvestVMs to enable the use of different instance types, to minimize costs while providing predictable execution times.

    \smartparagraph{Cloud configuration optimization.}
    Several research works target automatic cloud configuration optimization. Cherrypick~\cite{alipourfard17cherrypick}, Arrow~\cite{hsu2018arrow}, Scout~\cite{hsu2018scout}, Micky~\cite{micky}, Vanir~\cite{bilal2020vanir} and Lynceus~\cite{lynceus} perform cloud configuration optimization for distributed data analytics frameworks such as Spark and Hadoop. Ernest~\cite{venkataraman2016ernest} creates an analytical model for Spark applications and uses that to optimize cloud configurations. PARIS~\cite{yadwadkar2017selecting} uses historical data and machine learning to quickly choose cloud configurations for tasks that run on single VM instances. Our work is not aimed at finding the best optimization algorithm for automatic resource allocation for serverless functions; instead, we deal with design space questions to show the potential opportunities and how they can be utilized. We  used the BO variants that were part of the discussion in some of these works, but this is orthogonal to our main contribution, since other black-box optimization methods in prior works can replace the BO variants. In particular, if an analytical performance model of a serverless function can be created, such a model can be used instead of black-box performance modeling techniques to potentially lower the prediction error and speed up convergence.

    \smartparagraph{Resource allocation in data centers.}
    Paragon~\cite{delimitrou2013paragon} and its follow-up Quasar~\cite{delimitrou2014quasar} propose heterogeneity and interference-aware schedulers for data center workloads. They use collaborative filtering to classify an unknown incoming job to assign resources to it. Similarly, DejaVu~\cite{vasic2012dejavu} also tackles the problem of allocating resources to workloads in a data center, but uses clustering instead of collaborative filtering. Despite tackling a different problem of data center scheduling, we note that their use of collaborative filtering and clustering could also replace the black-box optimization methods we have discussed, provided that data on the performance of a representative set of benchmarking applications is available.

\smartparagraph{Multi-objective optimization.} Our work builds on the techniques from research area of multi-objective optimization. While we reuse a set of specific methods from this area, there are several other multi-objective optimization schemes \cite{arora2004introtoopt, marler2004mosurvey} that can be potentially used to provide a cost-performance trade-off for serverless users.

\section{Conclusion}
In this work, we demonstrated the benefits of decoupling memory and CPU resource allocations and using different instance types for serverless functions. Using the performance data we have collected, we established a potential to improve execution cost and execution time by up to 50\% and 40\% by decoupling CPU and memory resource allocations and utilizing different VM types for serverless functions. 

One way to deal with the increased number of resource allocation choices is by using black-box optimization methods. Our results showed that sampling-based and model-based black-box methods could find configurations that are within 10-20\% of the performance of the best configuration within 20 optimization trials. However, model-based methods are the obvious choice if predictions for untested configurations are needed. We found BO with GP to be the best model-based method in terms of reaching the best configuration within a limited number of optimization trials and providing a much lower prediction error than other BO variants we tested. Additionally, our evaluation showed that good configurations for one input sample are generally good for others: performing an input-specific optimization leads to less than 20\% improvement in execution and execution cost.

We outlined three types of interfaces to the user and their underlying optimization methodology to allow the user to select different cost-performance trade-off points. Lastly, we showed that, even in the presence of prediction error, we could utilize our performance models to enable the use of different instance types while providing predictable performance. This is key to allowing a cloud provider to use the idle resources of non-optimal instance types while minimizing the performance variation for end-user.  

\bibliographystyle{abbrv}
\balance
\bibliography{references}

\end{document}